\documentclass[12pt]{iopart}
\pdfoutput=1
\usepackage{graphicx}

\begin{document}

\title[An experimental test of gravity at high energy]{An experimental test of gravity at high energy}

\author{J-F. Glicenstein}

\address{IRFU, CEA, Universit\'e de Paris-Saclay, F91191 Gif-sur-Yvette}
\ead{glicens@cea.fr}
\vspace{10pt}
\begin{indented}
\item[]August 2018
\end{indented}

\begin{abstract}
Gravitational lensing of very high energy photons has recently been observed in the JVAS B0218+357 strong lensing system. 
This observation opens the possibility of performing a test of 
gravity at high energy
by comparing the difference in  propagation time of high energy photons over different travel paths. The time delay is computed in the framework 
of a LIV (Lorentz Invariance Violation) extension of the equations of motion of photons in the field of a massive object. 
However, the method obtained can be transposed to other models of gravity at high energy.   
 The potential for constraining high energy gravity with future observations of JVAS B0218+357 is discussed.  
 The bounds on  the LIV energy scale will not be competitive with other  astrophysical bounds such as those coming from 
 AGN and GRB flares.  However, these bounds are free of any assumption on the emission process. 
\end{abstract}

%
\vspace{2pc}
\noindent{\it Keywords}: quantum gravity phenomenology, gravitational lensing, gamma ray experiments
%

\submitto{\JCAP}
%
%
%

\section{Introduction}
Observing gravitational interactions of high energy particles is especially challenging, since  these observations  must involve at the same time large mass objects such as 
galaxies acting as lenses and high energy cosmic rays. High and very high energy photons have been observed only in two strong lensing systems:
PKS 1830-211 and  JVAS B0218+357. The evidence for lensing at high energy in PKS 1830-211 has been disputed \cite{2011A&A...528L...3B,2015ApJ...799..143A} and no lensing signal has been observed at very high energy \cite{2015ICRC...34..737G}. The situation is different for 
JVAS B0218+357, for which lensing of high energy \cite{2014ApJ...782L..14C} and very high energy \cite{2016A&A...595A..98A} photons has been observed. The high energy time delay between the two compact components of JVAS B0218+357 has been measured and is compatible with radio measurements \cite{2018MNRAS.476.5393B}. 

There is no universally accepted model for gravitational interactions at high energy and very few experimental constraints. 
One of the most popular tests of models consists in searching for energy dependent photon propagation from 
distant sources. These searches often assume a violation of Lorentz invariance and the results are expressed as bounds on the Lorentz Invariance Violation (LIV) energy scale. Very constraining results on the LIV energy scale have been obtained by looking at a variety of sources such as AGN or GRB (see e.g. \cite{2019ApJ...870...93A} and references therein).  
Photon deflection by massive objects and gravitational time delays have also been investigated in the context of the Standard Model Extension (SME)  \cite{2011PhRvD..84h5025T}, \cite{2009PhRvD..80d4004B} and ranbow gravity \cite{2012IJMPD..2150007G,2017PhLB..772..152D}. These computations generally use the Schwarzschild metric and they give constraints based on solar tests.
 An early study of lensing with LIV was published by Biesiada and Pi\'orkowska \cite{2009MNRAS.396..946B}. These authors  extrapolate the formula by Jacob and Piran \cite{2008JCAP...01..031J} to lensing delays by introducing a geometric "LIV comoving distance". They estimated that the observation of 20 TeV photons in the HST 14176+5226 lensing system would provide delays of tens of nanoseconds if the LIV scale is the Planck scale.  
 The formula by Jacob and Piran used in reference \cite{2009MNRAS.396..946B} is suited to the propagation of light in a cosmological context with some caveats \cite{2015PhRvD..92l4042R}. But is is not
 adapted to lensing studies, since the lensing time delay comes mostly from the local distorsion of photon trajectories in the vicinity of the lens.  
Photon trajectories and time delays could be further affected by a possible energy dependence of the Schwarzschild metric or Newton's constant \cite{2004CQGra..21.1725M}. 
 
In the first section of the paper, the lensing time delay and gravitational deflection of high energy photons is recalculated by an Hamiltonian method. The Hamiltonian used is a special case of the general LIV Hamiltonians with spherical symmetry given by Barcaroli et al  \cite{2017PhRvD..96h4010B}.  The formula obtained is then applied in the next section to future observations of JVAS B0218+357.  The sensitivity to the LIV energy scale is discussed and equivalent limits in the case of the rainbow metric are given.

\section{Single particle Hamiltonian with LIV in the field of a massive object}\label{sec:theory}

The Hamiltonian of particle motion in a spherical potential with Lorentz Invariance Violation (LIV) has the form

\begin{equation}
H	= -\frac{p_t^2}{f(r)}
	+ (1+ {\epsilon l_{P}}( \frac{p_t}{\sqrt{f(r)}}))\bigg[  f(r) p_r^2 + \frac{1}{r^2} {w}^2\bigg]\,.
\label{eq:HSCh}
\end{equation}

where $w^2 = p_{\theta}^2+\frac{p_{\phi}^2}{\sin^2{\theta}}, $ $p_t, p_r, p_{\theta}, p_{\phi}$ are the usual momenta conjugate to $t, r, \theta, \phi$ and $\epsilon =  \pm 1$ depending
whether the particle motion is super- or supraluminic. It depends on the unknown energy scale of LIV, which is denoted by $l_P$ and is of the order of the inverse of the Planck mass.
This Hamiltonian is a special case of a more general Lorentz Invariance Violating Hamiltonian given by Barcaroli et al \cite{2017PhRvD..96h4010B} which depends only on $l_{P}p_{t}$ 
and not on $l_P p_r.$  

In the Schwarzschild metric, one has $f(r)= 1-\frac{r}{r_S}$ where $r_S = \frac{2GM_L}{r}$ is the Schwarzschild radius and $M_L$ is the mass of the lensing object.  
The light deflection angle and time delay with LIV in the Schwarzschild metric are derived in the Appendix. 

In this paper, we are interested in real lenses which are not well described by Schwarzschild lenses. 
For a more general astrophysical lens, one has
$f(r) = 1+2U(r)$ with $U(r) \ll 1.$ The lens is still assumed to be spherically symmetric. 
To deal with these lenses, 
it is convenient to transform the Hamiltonian with a rescaling of the $r$ variable and use isotropic coordinates. The Hamiltonian in isotropic coordinates is:

\begin{equation}
H	= -\frac{p_t^2}{f(r)}  
+ (1+ {\epsilon l_{P}}( \frac{p_t}{\sqrt{f(r)}}))f(r) \vec{P}^2.
\end{equation}
with $\vec{P}^2 =\sum_1^3 {p_{\alpha}}^2.$

Hamilton's equation are obtained by derivating w.r.t the affine parameter $\lambda$ giving  
\begin{eqnarray*}
\frac{dp_{t}}{d\lambda}&= -\frac{\partial H}{\partial t}  = 0\\
\frac{dt}{d \lambda} &= \frac{\partial H}{\partial p_{t}} = -2\frac{p_{t}}{f(r)}+ (\frac{\epsilon l_P}{\sqrt{f(r)}}))\vec{P}^2 \\
\frac{dx^{\alpha}}{d\lambda} &= \frac{\partial H}{\partial p_{\alpha}} = 2f(r) p^{\alpha} (1+ {\epsilon l_{P}}( \frac{p_t}{\sqrt{f(r)}})) \\
\frac{dp_{\alpha}}{d\lambda} &=  -\frac{\partial H}{\partial x^{\alpha}}= -\partial_{\alpha} f(r) \left( \frac{p_t^2}{f(r)^2} + \vec{P}^2(1+{\epsilon l_{P}}( \frac{p_t}{2\sqrt{f(r)}}))  \right) 
\end{eqnarray*}

The mass constraint relevant to photon motion is $H = 0,$ which gives a relation between $P,$ $f(r)$ and $p_t$
 \begin{equation}
\frac{p_t}{P} = f \sqrt{ (1+ \frac{\epsilon l_{P} p_t}{\sqrt{f}}) } \simeq  f (1+ \frac{\epsilon l_{P} p_t}{2\sqrt{f}})  
\end{equation}

Hamilton's equation can be transformed using the euclidian line element 
$dl^2= \sum_1^3 {dx_{\alpha}}^2$ to give the evolution of $t$ and $p_{\alpha}$ with $l.$

Taking 
\begin{equation*}
\frac{dl}{d\lambda} =  -2f(r) P (1+ {\epsilon l_{P}}( \frac{p_t}{\sqrt{f(r)}})) \label{eq:defl} 
\end{equation*}

one has
 \begin{eqnarray}
\frac{dp_{\alpha}}{dl} & 
 = \partial_{\alpha} f(r) P \frac{\left( 2 +   {\epsilon l_{P}}( \frac{3p_t}{2\sqrt{f(r)}})   \right)}{2 f(r)(1+ {\epsilon l_{P}}( \frac{p_t}{\sqrt{f(r)}}))} \label{eq:momentum}\\
\frac{dt}{dl} &= \frac{p_{t}}{f^2(r)P}(1- \frac{\epsilon l_{P} p_t}{\sqrt{f}}) - (\frac{\epsilon l_PP}{2\sqrt{f(r)}})(1- \frac{\epsilon l_{P} p_t}{\sqrt{f}}) \label{eq:time}
\end{eqnarray}

In equation \ref{eq:momentum} $\partial_{\alpha} f(r) = 2\partial_{\alpha} U(r) $ is a first order quantity in $U.$  To first order in $U$, the evolution of  
momenta is thus given by

\begin{equation}
\frac{dp_{\alpha}}{dl} 
 = 2\partial_{\alpha} U(r) P \left( 1 -   {\epsilon l_{P}}( \frac{p_t}{4})   \right) \label{eq:momentum2}
 \end{equation}

The right hand side of equation \ref{eq:time} is the sum of two terms. The first one is 

 \begin{eqnarray*}
I_1 &= \frac{p_{t}}{f^2(r)P}(1- \frac{\epsilon l_{P} p_t}{\sqrt{f}}) =\frac{1}{f} \left( 1 - \frac{\epsilon l_{P} p_t}{2\sqrt{f}} \right)      \\
&=  (1-2U)(1-(1-U) \frac{\epsilon l_{P} p_t}{2}) \\
& = (1-\frac{\epsilon l_{P} p_t}{2}) -2U(1 -\frac{3}{4}\epsilon l_{P} p_t) +O(U^2) 
\end{eqnarray*}
and the second is
\begin{eqnarray*}
I_2 &= -(\frac{\epsilon l_PP}{2f(r)^{1/2}})(1- \frac{\epsilon l_{P} p_t}{\sqrt{f}}) = -(\frac{\epsilon l_Pp_t}{2 f(r)^{3/2}})(1- \frac{\epsilon l_{P} p_t}{\sqrt{f}})^2 \\
& = -(\frac{\epsilon l_Pp_t}{2 f(r)^{3/2}}) + O({l_{P}}^2 {p_t}^2) \\
& = -(\frac{\epsilon l_Pp_t}{2})(1 - 3U)
\end{eqnarray*}

Summing the 2 contributions:
\begin{equation}
\frac{dt}{dl} = (I_1+ I_2) = (1 -  \epsilon l_{P} p_t) -2U(1 -\frac{3}{2}\epsilon l_{P} p_t)
\label{eq:traveltime1}
\end{equation}

The rest of this section follows closely the formalism described in \cite{2017ApJ...850..102G}.
The thin lens approximation is used, photons
 move on a straight lines until they
get deflected toward the observer.  
The distance to the lens, located at redshift $z_L$ is $D_{OL}$, the distance from the lens to the observer is $D_{LS}$.  
Coordinates are
taken in the plane perpendicular to the line of sight (the Òlens
planeÓ). The projected source position on the lens plane is at $\eta$,
and the impact parameter of the particle trajectory at $\zeta.$

A popular lens model is the Singular Isothermal Lens (SIL) lens model, characterized by $U(r) = 2\sigma_v^2 \ln{r}.$  $\sigma_v$ is the velocity dispersion of stars in the lens galaxy.  $\sigma_v^2$ is proportional to Newton's constant G.  Using equation \ref{eq:traveltime1} and following the steps outlined in \cite{2017ApJ...850..102G}, the travel time from source to observer in a SIL model is 
 \begin{equation}
\label{eq:TimeSIL}
    T_{OS}  = (z_{L}+1) \left( \frac{1}{2} (1 -  \epsilon l_{P} p_t) (\frac{1}{D_{OL}} + \frac{1}{D_{LS}})(\zeta-\eta)^2-2\pi\sigma_v^2(1 -\frac{3}{2}\epsilon l_{P} p_t) |\zeta|+T_0 \right)
 \end{equation}

The absolute value of the deflection angle is obtained from equation \ref{eq:momentum2} (see \cite{2017ApJ...850..102G} for details).
\begin{equation}
 \alpha = 2\pi\sigma_v^2 \left( 1 -   {\epsilon l_{P}}( \frac{p_t}{4})   \right)\mbox{sgn}(\zeta). 
\label{eq:alphaSIL}
 \end{equation}
 
 The lens equation is:
 \begin{equation}
\zeta-\eta = \mbox{sgn}(\zeta) l_{E},
\label{eq:SILPS}
\end{equation}
with the Einstein length $l_{E}$ defined by
  \begin{equation}
\label{eq:EinsteinLength}
  l_{E} = \frac{4\pi \sigma_v^2 \left( 1 -   {\epsilon l_{P}}( \frac{p_t}{4})   \right) D_{OL}D_{LS} }{(D_{OL}+D_{LS})}
 \end{equation}
 Equation \ref{eq:SILPS} has 2 solutions for $|\eta| \le l_{E}.$ These solutions are
\begin{eqnarray}
\zeta_{+} & = l_{E} + \eta \label{eq:combinedsol2.5}  \\
\zeta_{-} & = \eta - l_{E} \label{eq:combinedsol2.6}
\end{eqnarray}
Since $(\zeta_{+}-\eta)^2 = (\zeta_{-}-\eta)^2 = l_{E}^2,$  only the second term in equation  
\ref{eq:TimeSIL} contributes to the time delay between images. This term depends on 
\begin{equation*}
|\zeta_{+}| - |\zeta_{-}| = 2\eta.
\end{equation*}

The time delay between the $\zeta_{\pm}$ images is thus:
\begin{equation}
\label{eq:TimeSIL2}
    \Delta T = -4 (z_{L}+1) \pi\sigma_v^2(1 -\frac{3}{2}\epsilon l_{P} p_t)\eta = \Delta T (p_t =0) (1 -\frac{3}{2}\epsilon l_{P} p_t)
 \end{equation}

 The time difference in equation \ref{eq:TimeSIL2} is somewhat similar (up to multiplicative terms of order 1) to the lensing time delay quoted in  
 \cite{2009MNRAS.396..946B}, but is obtained by a completely different method.  The expression for the Einstein length (equation  \ref{eq:EinsteinLength})
 differs from that obtained in  \cite{2009MNRAS.396..946B}. 
 
 Since $\Delta T$ scales linearly with $\sigma_v^2,$ hence with $G,$ an equation similar to equation \ref{eq:TimeSIL2} would be obtained in gravity models 
 with an energy-dependent $G,$ for instance a lensing gravity rainbow metric  expressed with energy dependent coordinates (equation 43 of reference \cite{2004CQGra..21.1725M}).  
Introducing the "rainbow function" $g,$ one has
\begin{equation}
\label{eq:TimeSIL2}
    \Delta T =  \Delta T (p_t =0)  \frac{G(p_t)}{G(p_t=0)} =\frac{\Delta T (p_t =0)} { g(p_t)}.
 \end{equation}
 The $g$ function can be parametrized as $$g(E) = \frac{1}{(1-\alpha \frac{E}{M_P})},$$ where $M_P = 1.22\ 10^{19}$ GeV is the Planck mass.  

\section{Application to high energy AGN flares}

\begin{figure}[h]
\centering
\includegraphics[height=6cm]{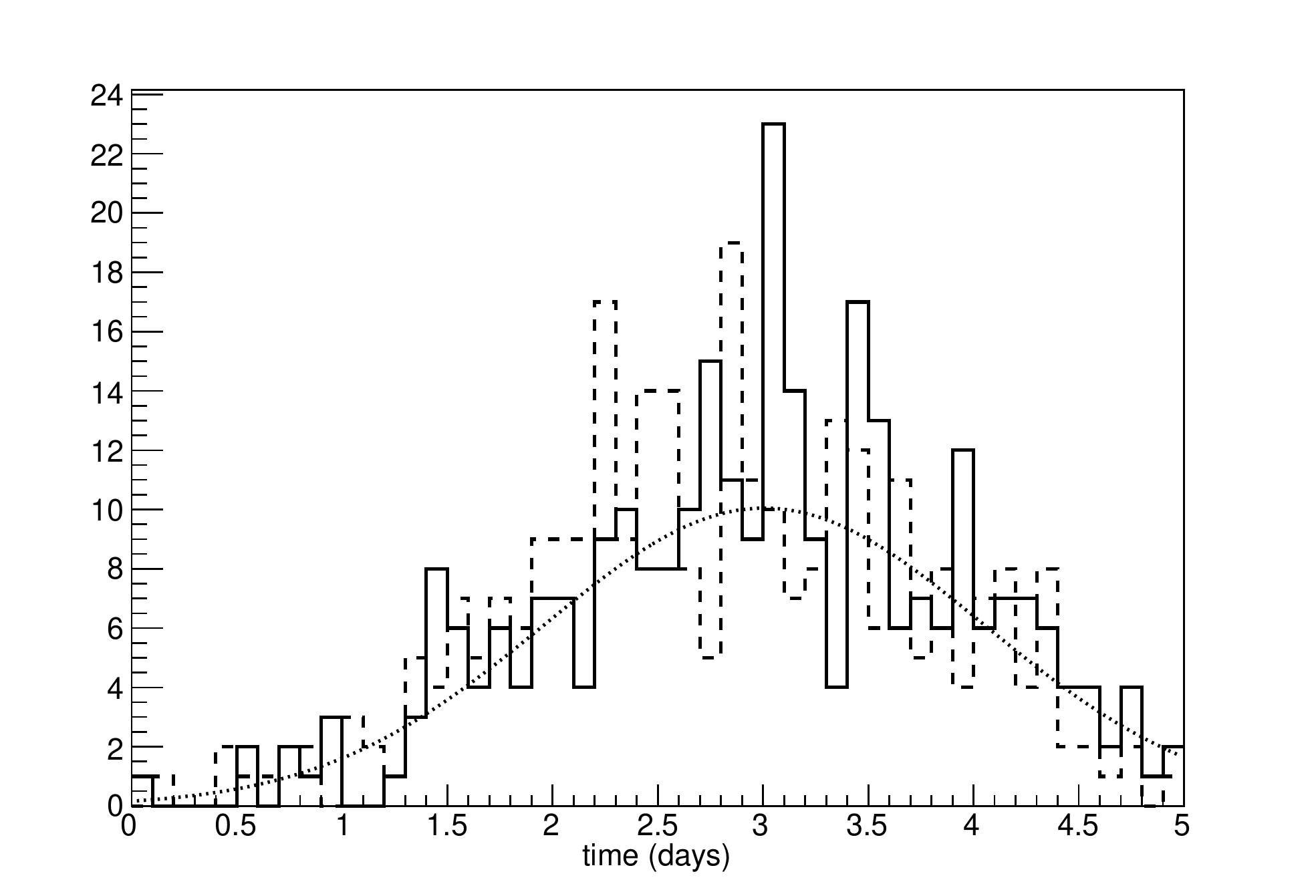}
\caption{Number of events versus date in a simulated lensing event. The original flare (solid line) has a gaussian time profile (dotted line). Photons from the delayed flare (dashed line) have an energy-dependent delay 
described by equation \ref{eq:TimeSIL2} with $l_P = 3.3\ 10^{-4} \mathrm{GeV}^{-1}.$ The constant part of the delay has been subtracted for clarity. 
}
\label{fig:observation}
\end{figure}

\begin{figure}[h]
\centering
\includegraphics[height=6cm]{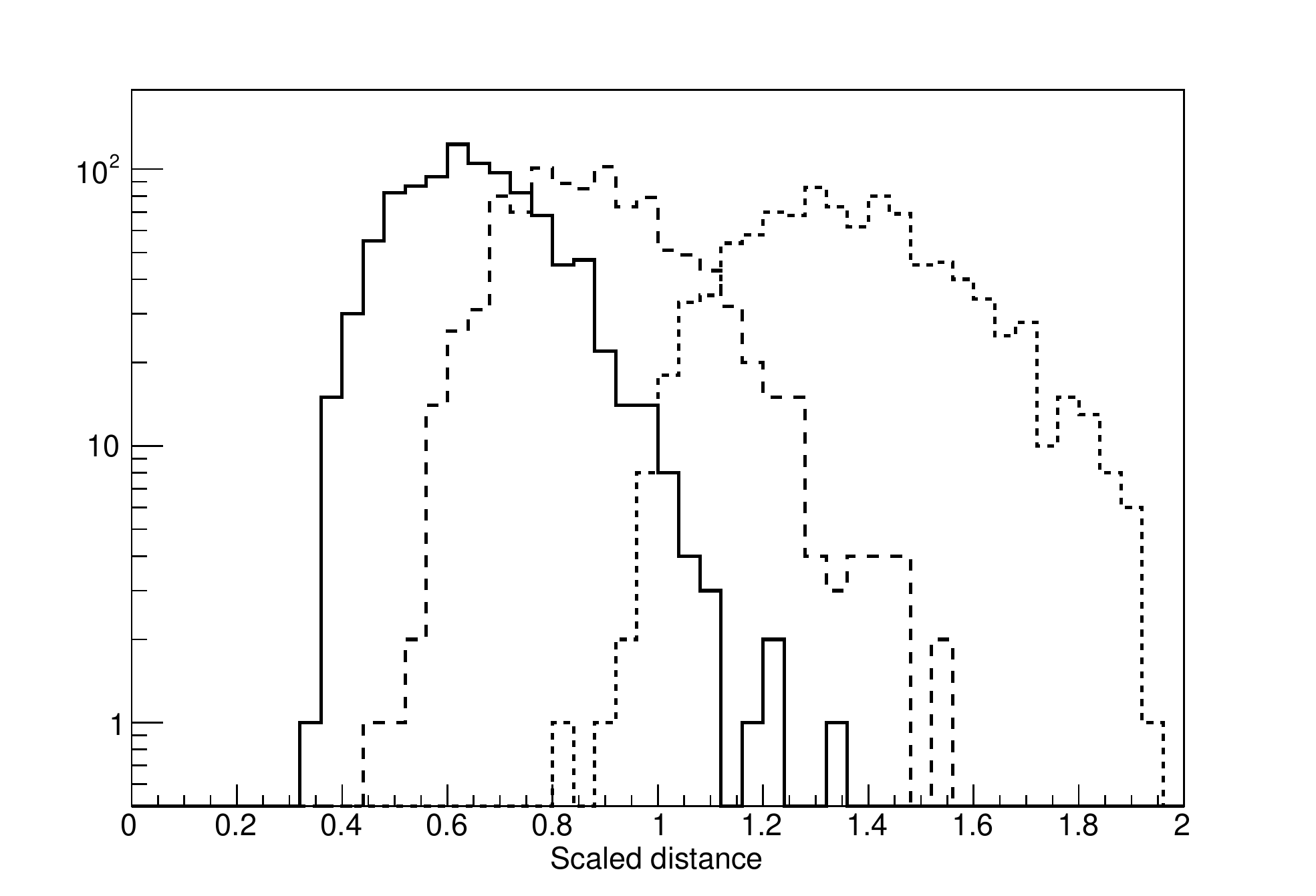}
\caption{Distribution of scaled Kolmogorov distance for simulated lensing events. The solid line shows the distribution with no LIV effect ($l_P = 0$), the long and short dashed lines respectively to 
$l_P = 1.3, \mathrm{and}\ 3.3\ 10^{-4} \mathrm{GeV}^{-1}.$
}
\label{fig:distance}
\end{figure}

As mentionned in the introduction, two lensed AGN have been observed in high and very high energy: PKS 1830-211 and  JVAS B0218+357. 
While the situation of PKS 1830-211 is still controversial, delayed lensing 
flares from JVAS B0218+357 have been detected in the high energy regime by the Fermi-LAT collaboration and in the very high energy regime by the MAGIC array of imaging Cerenkov telescopes. 
In addition, lens models of JVAS B0218+357 are available  \cite{2016ApJ...821...58B}. The JVAS B0218+357 lens is well approximated by a SIL model. 
More than 200 photons were detected in the range 65-175 GeV  during the MAGIC observation of a delayed flare of JVAS B0218+357 in 2014. The photon number density was well
described by a powerlaw $$\frac{dN}{dE} \propto E^{-\gamma}$$ with a photon index $\gamma = 3.8 \pm 0.6.$
In this paper, the sensitivity of future high energy detections to $l_P$is studied with a 
simulation.  The numbers used are slightly more optimistic than the MAGIC observations, since the forthcoming CTA observatory \cite{2017arXiv170907997C} will provide observations with a better 
sensitivity over a larger energy range. Thousand flaring events with $N_\gamma = 300$ detected photons both in the initial and delayed flare were simulated for a range of $l_P$ values. 
The energy range of the detected photons is 30 to 300 GeV. 
The simulated photons have a photon index of 3.8 and the initial flare has a gaussian luminosity profile in time (Fig. \ref{fig:observation}). The initial and delayed flare time profile are compared with a Kolmogorov-Smirnov test. When $l_P$  is large enough, the delayed flare has a distorted shape compared to the initial flare.  
The observable is the Kolmogov distance $\cal{D}.$ Figure \ref{fig:distance} shows the distribution of the scaled distance $\sqrt{\frac{N_{\gamma}}{2}} \cal{D}$ for values $l_P = 0, 1.3\ \mathrm{and}\ 3.3\ 10^{-4} \mathrm{GeV}^{-1}.$   A single observation could easily provide a limit on $l_P$ of several TeV$^{-1}$.   Simultaneous observations (e.g. with the Fermi-LAT instrument and ground based Imaging Cerenkov arrays)
would further improve the constraints. These limits on the LIV energy scale $\frac{1}{l_P}$ would be totally independent of the flare emission model, 
but more than 15 order of magnitude lower than the Planck mass.

As noted in the end of section \ref{sec:theory}, the constraint on $l_P$ is equivalent to a constraint on the $\alpha$ parameter of the "rainbow function" $g(E).$ 
Future observations of high energy photon lensing would give upper limits on $\alpha$ at the level of $10^{15},$ improving on the existing bounds \cite{2015EL....11020009F,2017PhRvD..96b4018B} by
several order of magnitudes.

\section{Conclusion}
In this paper, prospects for constraining the LIV energy scale with high energy observations of strong gravitational lenses have been discussed. Limits on the LIV energy scale of a few TeV can be reached. These limits
would give constraints much weaker than those obtained with AGN or GRB flares, but obtained with a cleaner setup, with no assumption about the emission process. On the other hand, 
limits on gravity rainbow models do not rely on 
the assumption of an energy dependent speed of light.  The observation of high energy photons lensed by systems such as JVAS B0218+357 could improve the existing limits on rainbow functions by several order of magnitude. 

\ack
I would like to thank Fran\c cois Brun for carefully reading the manuscript and for helpful suggestions.

\appendix
\setcounter{section}{1}
\section*{Appendix: Light deflection and time delay for a Schwarzschild lens with LIV}
This section contains the derivation of  the LIV extension of the usual deflection and time delay of light in the field of a massive compact astrophysical object described by a Schwarzschild 
metric. Derivating Hamiltonian in equation \ref{eq:HSCh} with respect to the affine parameter $\lambda,$ one obtains  a set of equations

\begin{eqnarray*}
\frac{dp_{t}}{d\lambda}&= -\frac{\partial H}{\partial t}  = 0\\
\frac{dt}{d \lambda} &= \frac{\partial H}{\partial p_{t}} = -2\frac{p_{t}}{f(r)}+ (\frac{\epsilon l_P}{\sqrt{f(r)}}))\bigg[  f(r) p_r^2 + \frac{1}{r^2} {p_{\theta}}^2\bigg] \\
\frac{dr}{d\lambda} &= \frac{\partial H}{\partial p_{r}} = 2f(r) p_{r} (1+ {\epsilon l_{P}}( \frac{p_t}{\sqrt{f(r)}})) \\
\frac{d\phi}{d\lambda} &=  \frac{\partial H}{\partial p_{\phi}}= 2\frac{p_{\phi}}{r^2 \sin{\theta}^2} (1+ {\epsilon l_{P}}( \frac{p_t}{\sqrt{f(r)}})) \\
\frac{d p_{\phi}}{d\lambda} &= -\frac{\partial H}{\partial \phi}= 0\\
\frac{d \theta}{d\lambda} &= \frac{\partial H}{\partial p_{\theta}}= \frac{2}{r^2}p_{\theta}(1+ {\epsilon l_{P}}( \frac{p_t}{\sqrt{f(r)}})) \\
\frac{d p_{\theta}}{d\lambda} &=  -\frac{\partial H}{\partial \theta} = \frac{d}{d\theta}(\frac{1}{2r^2\sin{\theta}^2}) p_{\phi}^2(1+ {\epsilon l_{P}}( \frac{p_t}{\sqrt{f(r)}}))  \\
\frac{d p_{r}}{d \lambda} &= -\frac{\partial H}{\partial r} = +\frac{f'(r)}{2f(r)^2}p_{t}^2 + \frac{1}{2}f'(r)p_{r}^2-\frac{1}{r^3}(p_{\theta}^2 +\frac{1}{\sin{\theta}^2} p_{\phi}^2)
\end{eqnarray*}

In these equations, $f(r)$ is given by $f(r)= 1-\frac{r}{r_S}$ where $r_S = \frac{2GM_L}{r}$ is the Schwarzschild radius and $M_L$ is the mass of the lensing object.  
$p_t, p_{\phi}$ are constants of motion. If in addition $\phi$ is constant, then $p_\theta$ is also a constant of motion.
The $\lambda$ affine parameter is eliminated and the remaining variables $t, \theta$ and $p_r$ are written as functions of $r.$
The $p_r$ variable can be further eliminated by the mass constraint $H=0,$ which gives

\begin{equation}
f(r) p_{r} = \sqrt{  \frac{p_t^2}{(1+ {\epsilon l_{P}}( \frac{p_t}{\sqrt{f(r)}}))} -  \frac{f(r)}{r^2} {p_{\theta}}^2}  
\end{equation}
Changing variable to $u=\frac{1}{r},$ one obtains 
\begin{eqnarray}
\frac{dt}{du} &=  \frac{p_{t}(1- (\frac{\epsilon l_P p_t}{2\sqrt{f(u)}(1+ {\epsilon l_{P}}( \frac{p_t}{\sqrt{f(u)}}))}))}{u^2f(u)^2 p_{r} (1+ {\epsilon l_{P}}( \frac{p_t}{\sqrt{f(u)}}))} \\
& \simeq  \frac{p_{t}(1- (\frac{3\epsilon l_P p_t}{2\sqrt{f(u)}}))}{u^2f(u)^2 p_{r} } \label{eq:dtdu2} \\
\frac{d \theta}{du} &=  -\frac{p_{\theta}}{\sqrt{  \frac{p_t^2}{(1+ {\epsilon l_{P}}( \frac{p_t}{\sqrt{f(u)}}))} -  f(u)u^2 {p_{\theta}}^2} }
\end{eqnarray}

The calculation follows the steps outlined in the book by Wald \cite{1984ucp..book.....W}.
The impact parameter is $\beta'$ defined by
 \begin{equation}
\frac{1}{{\beta'}^2} =  \frac{p_t^2}{p_\theta^2(1+ {\epsilon l_{P}}p_t)} 
\end{equation}

It is related to the solution $u_{0}$ of equation
\begin{equation}
\frac{p_t^2}{(1+ {\epsilon l_{P}}( \frac{p_t}{\sqrt{f(u)}}))} -  f(u)u^2 {p_{\theta}}^2=0 
\end{equation}
 by 
\begin{eqnarray}
\frac{1}{\beta'} = u_0 +\frac{r_S u_0^2(1/2 {\epsilon l_{P}}p_t -1)}{2} + O(r_S^2) \label{eq:betapvsrs}
\end{eqnarray}

Keeping only the lowest order in $r_S:$
\begin{eqnarray}
\frac{d \theta}{du} &=-\frac{1}{\sqrt{u_0^2 -u^2 + r_S(u^3-u_0^3) +\frac{\epsilon l_P p_t r_S u_0^2 }{2} (u_0-u)}} \\
&\simeq - \left(  \frac{1}{\sqrt{u_0^2 -u^2}}-\frac{r_S}{2} \left(\frac{(u^3-u_0^3)}{{(u_0^2 -u^2)}^{3/2}} + \frac{\epsilon l_P p_t}{2}  \frac{u_0^2(u_0-u)}{{(u_0^2 -u^2)}^{3/2}}    \right)        \right)
\end{eqnarray}
Integrating between 0 and $u_0,$ the deflection angle between the source and the lens is
\begin{equation}
\theta = -\frac{\pi}{2} +\frac{r_S}{\beta'} (1 - \frac{\epsilon l_P p_t}{4})
\end{equation}
Taking into account the deflection of light between the lens and the observer, the total observed deflection is:
\begin{equation}
\delta \theta = \frac{2r_S}{\beta'} (1 - \frac{\epsilon l_P p_t}{4})
\end{equation}
which is the generalisation of the usual deflection angle formula to first order in $l_p p_t.$

The calculation of time delay proceeds along similar lines.
After a somewhat lengthy calculation, the time delay of a signal sent from Earth and reflected off a planet is found to be
\begin{eqnarray}
\Delta T 
 &= 2\bigg( (1-\epsilon l_P p_t) (\sqrt{D_{LS}^2-R_0^2}+ \sqrt{D_{OL}^2-R_0^2}) \nonumber \\
&+ r_S(1-\frac{3}{2}\epsilon l_P p_t ) \left( \ln{\frac{D_{LS}+\sqrt{D_{LS}^2-R_0^2}}{{R_0}}} 
+\ln{\frac{D_{OL}+\sqrt{D_{OL}^2-R_0^2}}{{R_0}}}\right) \nonumber \\ 
&+\frac{1}{2} \left(\sqrt{\frac{D_{LS}-R_0}{D_{LS}+R_0}}  + \sqrt{\frac{D_{OL}-R_0}{D_{OL}+R_0}} \right)  \bigg)     
\end{eqnarray}
where $R_0 = \frac{1}{u_0}.$\newline\newline

						\bibliographystyle{unsrt}
						\bibliography{LensingwithLIV}

\end{document}